\PassOptionsToPackage{unicode}{hyperref}
\PassOptionsToPackage{hyphens}{url}
\PassOptionsToPackage{dvipsnames,svgnames,x11names}{xcolor}
\documentclass[
]{article}
\usepackage{amsmath,amssymb}
\usepackage{geometry}[margins=1.5cm]
\usepackage{lmodern}
\usepackage{iftex}
\ifPDFTeX
  \usepackage[T1]{fontenc}
  \usepackage[utf8]{inputenc}
  \usepackage{textcomp} 
\else 
  \usepackage{unicode-math}
  \defaultfontfeatures{Scale=MatchLowercase}
  \defaultfontfeatures[\rmfamily]{Ligatures=TeX,Scale=1}
\fi
\IfFileExists{upquote.sty}{\usepackage{upquote}}{}
\IfFileExists{microtype.sty}{
  \usepackage[]{microtype}
  \UseMicrotypeSet[protrusion]{basicmath} 
}{}
\makeatletter
\@ifundefined{KOMAClassName}{
  \IfFileExists{parskip.sty}{%
    \usepackage{parskip}
  }{
    \setlength{\parindent}{0pt}
    \setlength{\parskip}{6pt plus 2pt minus 1pt}}
}{
  \KOMAoptions{parskip=half}}
\makeatother
\usepackage{xcolor}
\usepackage{longtable,booktabs,array}
\usepackage{calc} 
\usepackage{etoolbox}
\makeatletter
\patchcmd\longtable{\par}{\if@noskipsec\mbox{}\fi\par}{}{}
\makeatother
\IfFileExists{footnotehyper.sty}{\usepackage{footnotehyper}}{\usepackage{footnote}}
\makesavenoteenv{longtable}
\usepackage{graphicx}
\providecommand{\pandocbounded}[1]{#1}
\makeatletter
\def\maxwidth{\ifdim\Gin@nat@width>\linewidth\linewidth\else\Gin@nat@width\fi}
\def\maxheight{\ifdim\Gin@nat@height>\textheight\textheight\else\Gin@nat@height\fi}
\makeatother
\setkeys{Gin}{width=\maxwidth,height=\maxheight,keepaspectratio}
\makeatletter
\def\fps@figure{htbp}
\makeatother
\NewDocumentCommand\citeproctext{}{}
\NewDocumentCommand\citeproc{mm}{%
  \begingroup\def\citeproctext{#2}\cite{#1}\endgroup}
\makeatletter
 \let\@cite@ofmt\@firstofone
 \def\@biblabel#1{}
 \def\@cite#1#2{{#1\if@tempswa , #2\fi}}
\makeatother
\newlength{\cslhangindent}
\newlength{\csllabelwidth}
\newenvironment{CSLReferences}[2] 
 {\begin{list}{}{%
  \setlength{\itemindent}{0pt}
  \setlength{\leftmargin}{0pt}
  \setlength{\parsep}{0pt}
  \ifodd #1
   \setlength{\leftmargin}{\cslhangindent}
   \setlength{\itemindent}{-1\cslhangindent}
  \fi
  \setlength{\itemsep}{#2\baselineskip}}}
 {\end{list}}
\usepackage{calc}

\ifLuaTeX
\usepackage[bidi=basic]{babel}
\else
\usepackage[bidi=default]{babel}
\fi
\babelprovide[main,import]{american}

\def\languageshorthands#1{}
\ifLuaTeX
  \usepackage{selnolig}  
\fi
\IfFileExists{bookmark.sty}{\usepackage{bookmark}}{\usepackage{hyperref}}
\IfFileExists{xurl.sty}{\usepackage{xurl}}{} 
\hypersetup{
  pdftitle={hyprfine: simulating the 21-cm signal from the Dark Ages
through to the Epoch of Reionization on a GPU},
  pdfauthor={Harry T. J. Bevins},
  pdflang={en-US},
  colorlinks=true,
  linkcolor={Maroon},
  filecolor={Maroon},
  citecolor={Blue},
  urlcolor={Blue},
  pdfcreator={LaTeX via pandoc}}

\title{hyprfine: simulating the 21-cm signal from the Dark Ages through
to the Epoch of Reionization on a GPU}

\definecolor{c53baa1}{RGB}{83,186,161}
\definecolor{c202826}{RGB}{32,40,38}

\usepackage[affil-it]{authblk}
\usepackage{orcidlink}
\author[1,2%
  ]{Harry T. J. Bevins%
    \,\orcidlink{0000-0002-4367-3550}\,%
    }

\affil[1]{Department of Physics, Imperial College London, Blackett
Laboratory, Prince Consort Road, London SW7 2AZ, UK%
  }
\affil[2]{I-X Centre for AI in Science, Translation and Innovation Hub
(I-HUB), Imperial College London, White City Campus, 84 Wood Lane,
London W12 0BZ, UK%
  }
\date{01 October 2026}

\begin{document}
\maketitle

\section{Summary}\label{summary}

The field of 21-cm Cosmology aims to observe the evolution of the early
universe between redshifts of 1100 and 6 (corresponding to roughly
400,000 years after the big bang to 1 billion years) through the 21-cm
emission from neutral hydrogen. The signal arises from the spin flip
transition in neutral hydrogen and the relative number of atoms with
aligned and anti-aligned proton and electron spins is characterized by a
statistical temperature called the spin temperature. At various times
the spin temperature \(T_s\), is driven by interactions with Cosmic
Microwave Background (CMB) photons with a wavelength of 21-cm,
collisions between hydrogen atoms in the gas (\(z \approx 300 - 30\)),
light from the first stars (\(z \approx 30 -6\)) and X-ray emission from
exotic objects like X-ray binaries (\(z \approx 20 - 6\)). These
processes couple the spin temperature to the gas temperature \(T_k\)
which is cooler than the CMB \(T_\gamma\) and so the 21-cm signal is
seen in absorption against the radio background at different epochs.

\texttt{hyprfine} is an analytic simulation of the sky-averaged 21-cm
signal from \(z=1100 - 6\) written using JAX and Python for native GPU
capabilities. It models the average temperature of the 21-cm signal over
cosmic time given by

\[T_{21} = 54 (1 - x_e) \frac{(1 - Y_{\rm He})}{0.76} \frac{\Omega_b h^2}{0.02242} \sqrt{\frac{0.1424}{\Omega_m h^2}\frac{1+z}{40}}\bigg(1 - \frac{T_{\rm CMB}}{T_s}\bigg)\]

as a function of the \(\Lambda\)-CDM cosmology parameters and the
astrophysics of the first stars and galaxies. As far as we are aware,
the code is the first analytic GPU native simulation of the 21-cm
signal. It runs in a fraction of a second, parallelises efficiently
across a GPU and is differentiable through the Dark Ages
(\(z \geq 35\)). The code is publicly available on \href{https://www.github.com/htjb/hyprfine}{GitHub}.

\section{Statement of need}\label{statement-of-need}

A number of global or sky-averaged 21-cm signal experiments have
collected data in recent years or are currently collecting data (EDGES
Bowman et al. (\citeproc{ref-Bowman2018EDGES}{2018}); SARAS Singh et al.
(\citeproc{ref-Singh2022SARAS}{2022}), Bevins, Fialkov, et al.
(\citeproc{ref-Bevins2022SARAS}{2022}); REACH de Lera Acedo et al.
(\citeproc{ref-Acedo2022REACH}{2022})). In order to analyse this data
researchers rely on Bayesian inference techniques
(\citeproc{ref-Anstey2021BayesianForegrounds}{Anstey et al., 2021};
\citeproc{ref-Bevins2022SARAS2}{Bevins, de Lera Acedo, et al., 2022};
\citeproc{ref-Bevins2022SARAS}{Bevins, Fialkov, et al., 2022};
\citeproc{ref-Acedo2022REACH}{de Lera Acedo et al., 2022};
\citeproc{ref-Dhandha2025JWST21cm1}{Dhandha et al., 2025a},
\citeproc{ref-Dhandha2025JWST21cm2}{2025b};
\citeproc{ref-Pochinda2024PopIII}{Pochinda et al., 2024};
\citeproc{ref-Tutt2026GPU21cm}{Tutt, Sims, et al., 2026}) and neural
network emulators (\citeproc{ref-Bevins2021GLOBALEMU}{Bevins et al.,
2021}; \citeproc{ref-Breitman2024EMU21cm}{Breitman et al., 2024};
\citeproc{ref-Bye2022VAE21cm}{Bye et al., 2022};
\citeproc{ref-Cohen202021cmGEM}{Cohen et al., 2020};
\citeproc{ref-DorigoJones2024LSTM21cm}{Dorigo Jones et al., 2024},
\citeproc{ref-DorigoJones2025KAN21cm}{2025}) of complex semi-numerical
simulations (\citeproc{ref-Fialkov2012RelativeMotion}{Fialkov et al.,
2012}; \citeproc{ref-Mesinger2011FAST21cm}{Mesinger et al., 2011};
\citeproc{ref-Murray2020FAST21cm}{Murray et al., 2020};
\citeproc{ref-Visbal2012FirstStars}{Visbal et al., 2012}).

These semi-numerical simulations take order hours to run per parameter
set and in inference loops the model has to be called 100,000s to
millions of times. As a result emulators have become a crucial piece of
infrastructure in the field with the state-of-the-art emulators being
able to accurately recover the 21-cm signal in a fraction of a second.
However, emulators are inherently approximations of the underlying
physics model (\citeproc{ref-Bevins2025PosteriorRecovery}{Bevins et al.,
2025}) and \texttt{hyprfine} evaluates the model directly in tens of
milliseconds on a GPU, without training an emulator on the full signal.

\texttt{hyprfine} currently includes key effects like collisional
coupling, the Wouthuysen-Field and X-ray heating. Additional effects
such as Lyman-\(\alpha\) heating can easily be added and the code
developed while maintaining the parallelism afforded by modern GPU
architectures.

\section{State of the field}\label{state-of-the-field}

\begin{figure}
\centering
\pandocbounded{\includegraphics[keepaspectratio]{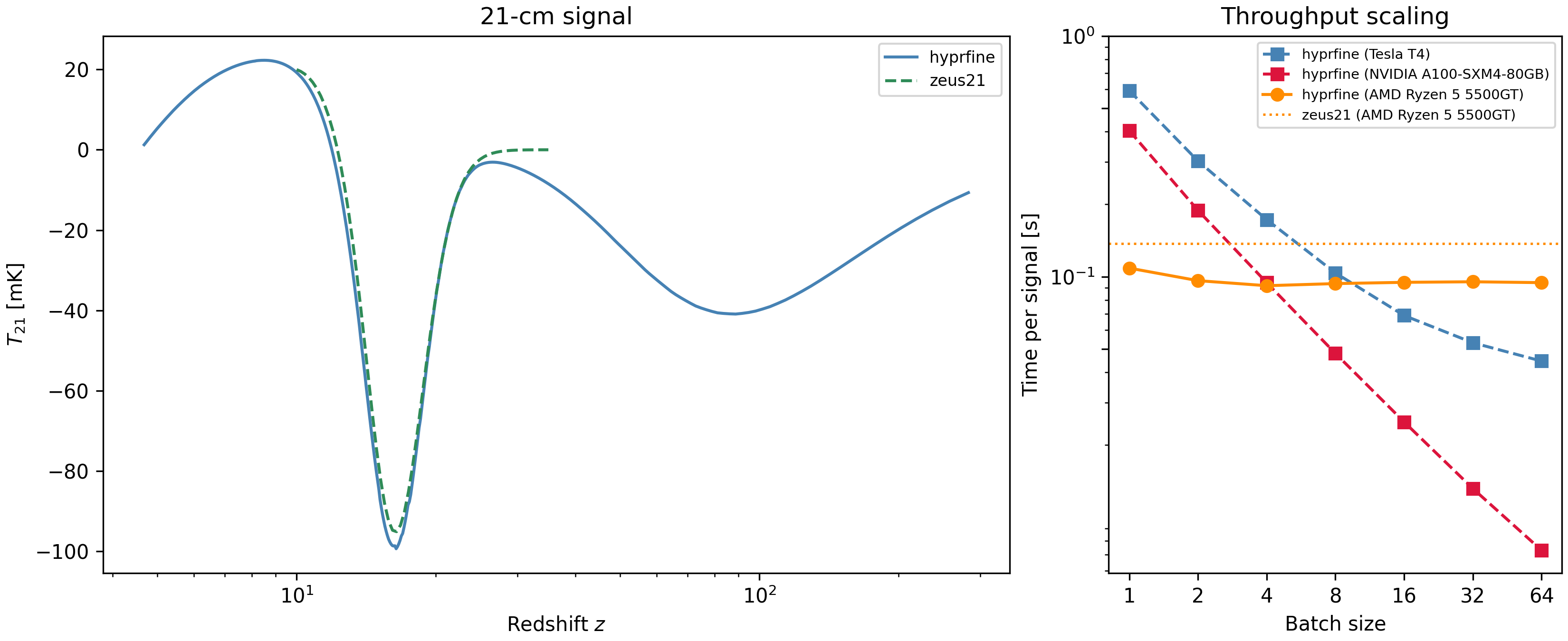}}
\caption{\textbf{Left:} An example 21-cm signal from hyprfine and zeus21
with approximately the same parameters. \textbf{Right:} The run time of
hyprfine per signal on an AMD Ryzen 5, a Tesla T4 and an A100 compared
to the value for zeus21. \label{fig:benchmark}}
\end{figure}

21-cm simulation codes fall broadly into three classes, numerical,
semi-numerical and analytic, which trade physical detail for speed
(\autoref{tab:codes}). Only a small number of publicly available
simulation codes make use of GPUs.

Numerical codes solve the radiative transfer directly. \(C^2\)-ray
(\citeproc{ref-Mellema2006C2ray}{Mellema et al., 2006}) is written in
Fortran90 and post-processes cosmological N-body simulations with
ray-tracing to compute the ionization field during the Epoch of
Reionization (\(z \approx 20 - 6\)), from which the 21-cm signal can be
derived. Its successor pyC\(^2\)ray
(\citeproc{ref-Hirling2024pyc2ray}{Hirling et al., 2024}) moves the
ray-tracing to C++ and CUDA with a Python interface. On an NVIDIA Tesla
P100 it post-processes a (349 Mpc)\(^3\) N-body simulation in
\(\approx 2.5\) GPU hours, compared with 13,824 core hours for the
original CPU code. Even on a GPU, simulations like these are far too
expensive for parameter inference.

Semi-numerical codes such as 21cmFAST
(\citeproc{ref-Mesinger2011FAST21cm}{Mesinger et al., 2011};
\citeproc{ref-Murray2020FAST21cm}{Murray et al., 2020}) and 21cmSPACE
(\citeproc{ref-Fialkov2012RelativeMotion}{Fialkov et al., 2012};
\citeproc{ref-Visbal2012FirstStars}{Visbal et al., 2012}) use
approximate methods such as excursion-set ionization and analytic
radiation fields to produce 21-cm boxes from which the global signal and
power spectrum are calculated. They take on the order of hours per
parameter set, which is why emulators of them are used for inference.

Analytic codes compute the sky-averaged quantities directly and are the
closest relatives of \texttt{hyprfine}. ARES
(\citeproc{ref-Mirocha2014ARES}{Mirocha, 2014}) models the global signal
from cosmic dawn onwards and treats the radiation background with
one-dimensional radiative transfer. Zeus21
(\citeproc{ref-Munoz2023Zeus21}{Muñoz, 2023}) computes the global signal
and power spectrum in \(\lesssim 1\) s by exploiting the approximately
exponential dependence of the star formation rate density on the density
field, and agrees with 21cmFAST to \(\sim 10\%\). ECHO21
(\citeproc{ref-Mittal2026ECHO21}{Mittal et al., 2026}) models the global
signal from the Dark Ages to the end of reionization in
\(\mathcal{O}(1)\) s, includes Lyman-\(\alpha\) heating and allows the
cosmological parameters to vary. All three run on CPUs with NumPy and
SciPy, and ECHO21 speeds up large model grids by spreading them across
CPU cores.

\texttt{hyprfine} follows these codes in its physics, for example in
using the star formation rate density model of Muñoz
(\citeproc{ref-Munoz2023Zeus21}{2023}), and \autoref{fig:benchmark}
compares it with Zeus21 for identical parameters. The difference is
where and how it runs. Making an existing code GPU-native and
differentiable would mean rewriting every integrator, interpolator and
ODE solver in JAX and replacing calls to external CPU codes (e.g.~CLASS
(\citeproc{ref-Blas2011CLASS2}{Blas et al., 2011}) in Zeus21,
recombination codes (\citeproc{ref-Lee2020HYREC2}{Lee \& Ali-Ha\"{\i}moud,
2020})). That amounts to a rewrite, so \texttt{hyprfine} was written as
a new, JAX-native package. It is the first analytic 21-cm code that runs
natively on a GPU, evaluates large batches of models in parallel and
provides gradients with respect to its parameters.

\pagebreak
\begin{longtable}[]{@{}
  >{\raggedright\arraybackslash}p{(\linewidth - 10\tabcolsep) * \real{0.1078}}
  >{\raggedright\arraybackslash}p{(\linewidth - 10\tabcolsep) * \real{0.1863}}
  >{\raggedright\arraybackslash}p{(\linewidth - 10\tabcolsep) * \real{0.2549}}
  >{\raggedright\arraybackslash}p{(\linewidth - 10\tabcolsep) * \real{0.1275}}
  >{\raggedright\arraybackslash}p{(\linewidth - 10\tabcolsep) * \real{0.0784}}
  >{\raggedright\arraybackslash}p{(\linewidth - 10\tabcolsep) * \real{0.2451}}@{}}
\caption{An inexhaustive list of publicly available 21-cm simulation
codes with their simulation type, outputs, approximate redshift range,
GPU support, and implementation language.
\label{tab:codes}}\tabularnewline
\toprule\noalign{}
\begin{minipage}[b]{\linewidth}\raggedright
Code
\end{minipage} & \begin{minipage}[b]{\linewidth}\raggedright
Simulation Type
\end{minipage} & \begin{minipage}[b]{\linewidth}\raggedright
Products
\end{minipage} & \begin{minipage}[b]{\linewidth}\raggedright
Approx. Redshift
\end{minipage} & \begin{minipage}[b]{\linewidth}\raggedright
GPU?
\end{minipage} & \begin{minipage}[b]{\linewidth}\raggedright
Programming Lang./Framework
\end{minipage} \\
\midrule\noalign{}
\endfirsthead
\toprule\noalign{}
\begin{minipage}[b]{\linewidth}\raggedright
Code
\end{minipage} & \begin{minipage}[b]{\linewidth}\raggedright
Simulation Type
\end{minipage} & \begin{minipage}[b]{\linewidth}\raggedright
Products
\end{minipage} & \begin{minipage}[b]{\linewidth}\raggedright
Approx. Redshift
\end{minipage} & \begin{minipage}[b]{\linewidth}\raggedright
GPU?
\end{minipage} & \begin{minipage}[b]{\linewidth}\raggedright
Programming Lang./Framework
\end{minipage} \\
\midrule\noalign{}
\endhead
\bottomrule\noalign{}
\endlastfoot
pyC\(^2\)ray & Numerical & Spatially resolved ionization field & 21 - 6
& Yes (partial) & Python, Fortran90, C++, CUDA \\
21cmFAST & Semi-numerical & 21-cm simulation box, summary statistics and
more & 50 - 6 & No & Python, C \\
Zeus21 & Analytic & 21-cm global signal, power spectrum, UVLF & 35 - 5 &
No & Python \\
ECHO21 & Analytic & 21-cm global signal & 1500 - 0 & No & Python \\
ARES & Semi-analytic/1D RT & 21-cm global signal and more & 35 - 5 & No
& Python \\
\texttt{hyprfine} & Analytic & 21-cm global signal & 1100 - 6 & Yes &
Python, JAX \\
\end{longtable}

\section{Software design}\label{software-design}

\texttt{hyprfine} is written in a functional style around JAX. Every
physical ingredient is a pure, just-in-time compiled function of
redshift and two parameter containers, \texttt{cosmology} and
\texttt{astrophysics}. These are Python named tuples and therefore JAX
pytrees, so the full simulation can be vectorised over parameter sets
with \texttt{jax.vmap}, distributed across GPU cores and differentiated
with \texttt{jax.grad} without any bespoke code. This choice favours
simple, composable functions over an object-oriented simulation class
with internal state, which would be harder to trace and compile.

The top-level \texttt{generate\_signal} function chains the physics
together. At \(z \geq 35\) the free electron fraction \(x_e\) and gas
temperature \(T_k\) are taken from a neural network emulator of the
recombination code HYREC-2, built with \texttt{astroemu}\footnote{\url{https://github.com/htjb/astroemu}}.
Wrapping the original C code would require leaving the GPU and would
break differentiability; the emulator keeps the whole pipeline on-device
at the cost of a small emulation error (\(\lesssim 3 \%\) in \(x_e\) and
\(\lesssim 1\%\) in \(T_k\)). Below \(z = 35\), \(T_k\), the background
ionization fraction and the filling factor of ionized bubbles are
evolved as a stiff system of ordinary differential equations, sourced by
X-ray heating and ionizing photons from the first galaxies, using the
implicit \texttt{Kvaerno5} solver from \texttt{diffrax}
(\citeproc{ref-Kidger2022NeuralDE}{Kidger, 2022}). Collisional and
Wouthuysen-Field coupling coefficients, the spin temperature and
\(T_{21}\) are then evaluated on the requested frequency grid. The code
runs in double precision throughout. Currently the Dark Ages portion of
the signal is differentiable, and extending gradients through the cosmic
dawn solve is ongoing work.

The code is organised into one module per physical process: cosmology,
the matter power spectrum and \(\sigma(M)\), the halo mass function and
star formation rate density, Lyman-\(\alpha\) emissivity, X-ray
emissivity, coupling coefficients and temperatures. Each module is
tested on its own. New physics such as Lyman-\(\alpha\) heating can
therefore be added as a new function and inserted into the ODE vector
field or the spin temperature calculation without changing the rest of
the code.

This modular structure is also designed to support a future extension to
the 21-cm power spectrum. The ingredients needed for a fluctuation model
are already general, differentiable functions of scale, mass and
redshift: the linear matter power spectrum, \(\sigma(M)\), growth
factors, the halo mass function, star formation efficiency and the
Lyman-\(\alpha\) and X-ray emissivities. These building blocks can be
reused directly by an analytic power spectrum model similar to that of
zeus21 or even a semi-numerical simulation, keeping the GPU parallelism
and differentiability of the global signal code.

\section{Research impact statement}\label{research-impact-statement}

An early version of \texttt{hyprfine} has already been used in published
research (\citeproc{ref-Bevins2026ApJL}{Bevins, 2026}), where it was
used to assess the constraining power of early and late time probes of
cosmology on the Dark Ages 21-cm signal.

More broadly, \texttt{hyprfine} targets a bottleneck in current global
21-cm analyses. Bayesian inference of data from experiments such as
EDGES, SARAS and REACH requires \(10^5\)--\(10^6\) model evaluations,
which has so far required neural network emulators trained on expensive
semi-numerical simulations. \texttt{hyprfine} evaluates the physical
model directly in a fraction of a second per signal on a GPU and can
evaluate large batches of parameter sets in parallel
(\autoref{fig:benchmark}), making inference with the physical model
itself practical. Its coverage from \(z=1100\) makes it suitable for
forecasting and analysis for proposed lunar Dark Ages experiments, and
its differentiability enables gradient-based samplers and Fisher
forecasts. As experiments move their inference pipelines to GPUs
(\citeproc{ref-Tutt2026GPU21cm}{Tutt, Sims, et al., 2026};
\citeproc{ref-Tutt2026Beams}{Tutt, Anstey, et al., 2026}) with
frameworks like BlackJAX (\citeproc{ref-cabezas2024blackjax}{Cabezas et
al., 2024}), \texttt{hyprfine} provides a fast and efficient GPU native
way to model the 21-cm signal. The code is released on PyPI under the
MIT licence with documentation, tutorials and a test suite, ready for
use and extension by the community.

\section{Conclusions}\label{conclusions}

\texttt{hyprfine} is in active development. We are currently assessing
the feasibility of adding additional physics into the model including
Lyman-\(\alpha\) heating (\citeproc{ref-Reis2021LyAlpha21cm}{Reis et
al., 2021}) and alternative dark matter models. Furthermore, we aim to
make the Cosmic Dawn component of the code differentiable and will
explore alternative ODE solvers like modax
(\citeproc{ref-Berry2026MODAX}{Berry et al., 2026}) and GRADSOLVE
(\citeproc{ref-SpurioMancini2026GRADSOLVE}{Spurio Mancini, 2026}). In
the long run we hope to see \texttt{hyprfine} evolve to include a GPU
compatible and differentiable semi-analytic model of the 21-cm signal
that can be rapidly evaluated either during inference or alternatively
to generate larger training data sets for more accurate emulation than
is possible with existing frameworks. A GPU compatible model of the
21-cm signal with a differentiable Dark Ages component is a timely
addition to the 21-cm inference landscape as researchers move their
pipelines on to the GPU (\citeproc{ref-Tutt2026GPU21cm}{Tutt, Sims, et
al., 2026}; \citeproc{ref-Tutt2026Beams}{Tutt, Anstey, et al., 2026})
and a number of experiments targeting observations of the dark ages
signal from the moon come online (\citeproc{ref-de2026cosmocube}{de Lera
Acedo et al., 2026}).

\texttt{hyprfine} offers a GPU compatible way to model the 21-cm signal
across a large period of cosmic history, is differentiable through the
dark ages and parallelisable.

\section{AI usage disclosure}\label{ai-usage-disclosure}

Anthropic's CLAUDE and CLAUDE Code were used in the development of
\texttt{hyprfine}, the writing of the documentation and the writing of
this paper. The models were used to write test suites, benchmarking
codes, helped debugging runtime issues and with software design choices.
All AI generated code and text were reviewed and validated by the author
who takes full responsibility for the code and paper. In particular
where AI models contributed to the coding of the physics modules, the
author rigorously checked the work against the literature.

\section{Acknowledgements}\label{acknowledgements}

HTJB acknowledges support from Imperial College London and Schmidt
Science through an Imperial College Research Fellowship. HTJB
acknowledges the use of resources provided by the Isambard 3 Tier-2 HPC
Facility (APP63640). Isambard 3 is hosted by the University of Bristol
and operated by the GW4 Alliance (https://gw4.ac.uk) and is funded by UK
Research and Innovation; and the Engineering and Physical Sciences
Research Council {[}EP/X039137/1{]}.

\section*{References}\label{references}
\addcontentsline{toc}{section}{References}

\protect\phantomsection\label{refs}


\begin{CSLReferences}{1}{0}
\bibitem[\citeproctext]{ref-Anstey2021BayesianForegrounds}
Anstey, D., de Lera Acedo, E., \& Handley, W. (2021). {A general
Bayesian framework for foreground modelling and chromaticity correction
for global 21 cm experiments}. \emph{MNRAS}, \emph{506}(2), 2041--2058.
\url{https://doi.org/10.1093/mnras/stab1765}

\bibitem[\citeproctext]{ref-Berry2026MODAX}
Berry, L., Handley, W., Hahn, O., \& Schöneberg, N. (2026).
{Accelerating massive ensembles of ordinary differential equations}.
\emph{arXiv e-Prints}, arXiv:2609.35661.
\url{https://arxiv.org/abs/2609.35661}

\bibitem[\citeproctext]{ref-Bevins2026ApJL}
Bevins, H. T. J. (2026). {The Magnitude of the Dark Ages 21 cm Signal in
the Context of Existing Early and Late Time Constraints on
{\(\Lambda\)}CDM}. \emph{ApJL}, \emph{997}(2), L35.
\url{https://doi.org/10.3847/2041-8213/ae356b}

\bibitem[\citeproctext]{ref-Bevins2022SARAS2}
Bevins, H. T. J., de Lera Acedo, E., Fialkov, A., Handley, W. J., Singh,
S., Subrahmanyan, R., \& Barkana, R. (2022). {A comprehensive Bayesian
reanalysis of the SARAS2 data from the epoch of reionization}.
\emph{MNRAS}, \emph{513}(3), 4507--4526.
\url{https://doi.org/10.1093/mnras/stac1158}

\bibitem[\citeproctext]{ref-Bevins2022SARAS}
Bevins, H. T. J., Fialkov, A., de Lera Acedo, E., Handley, W. J., Singh,
S., Subrahmanyan, R., \& Barkana, R. (2022). {Astrophysical constraints
from the SARAS 3 non-detection of the cosmic dawn sky-averaged 21-cm
signal}. \emph{Nature Astronomy}, \emph{6}, 1473--1483.
\url{https://doi.org/10.1038/s41550-022-01825-6}

\bibitem[\citeproctext]{ref-Bevins2025PosteriorRecovery}
Bevins, H. T. J., Gessey-Jones, T., \& Handley, W. J. (2025). {On the
accuracy of posterior recovery with neural network emulators}.
\emph{MNRAS}, \emph{544}(1), 375--390.
\url{https://doi.org/10.1093/mnras/staf1590}

\bibitem[\citeproctext]{ref-Bevins2021GLOBALEMU}
Bevins, H. T. J., Handley, W. J., Fialkov, A., de Lera Acedo, E., \&
Javid, K. (2021). {GLOBALEMU: a novel and robust approach for emulating
the sky-averaged 21-cm signal from the cosmic dawn and epoch of
reionization}. \emph{MNRAS}, \emph{508}(2), 2923--2936.
\url{https://doi.org/10.1093/mnras/stab2737}

\bibitem[\citeproctext]{ref-Blas2011CLASS2}
Blas, D., Lesgourgues, J., \& Tram, T. (2011). {The Cosmic Linear
Anisotropy Solving System (CLASS). Part II: Approximation schemes}.
\emph{JCAP}, \emph{2011}(7), 034.
\url{https://doi.org/10.1088/1475-7516/2011/07/034}

\bibitem[\citeproctext]{ref-Bowman2018EDGES}
Bowman, J. D., Rogers, A. E. E., Monsalve, R. A., Mozdzen, T. J., \&
Mahesh, N. (2018). {An absorption profile centred at 78 megahertz in the
sky-averaged spectrum}. \emph{Nature}, \emph{555}(7694), 67--70.
\url{https://doi.org/10.1038/nature25792}

\bibitem[\citeproctext]{ref-Breitman2024EMU21cm}
Breitman, D., Mesinger, A., Murray, S. G., Prelogović, D., Qin, Y., \&
Trotta, R. (2024). {21CMEMU: an emulator of 21CMFAST summary
observables}. \emph{MNRAS}, \emph{527}(4), 9833--9852.
\url{https://doi.org/10.1093/mnras/stad3849}

\bibitem[\citeproctext]{ref-Bye2022VAE21cm}
Bye, C. H., Portillo, S. K. N., \& Fialkov, A. (2022). {21cmVAE: A Very
Accurate Emulator of the 21 cm Global Signal}. \emph{ApJ},
\emph{930}(1), 79. \url{https://doi.org/10.3847/1538-4357/ac6424}

\bibitem[\citeproctext]{ref-cabezas2024blackjax}
Cabezas, A., Corenflos, A., Lao, J., \& Louf, R. (2024). \emph{BlackJAX:
Composable {B}ayesian inference in {JAX}}.
\url{https://arxiv.org/abs/2402.10797}

\bibitem[\citeproctext]{ref-Cohen202021cmGEM}
Cohen, A., Fialkov, A., Barkana, R., \& Monsalve, R. A. (2020).
{Emulating the global 21-cm signal from Cosmic Dawn and Reionization}.
\emph{MNRAS}, \emph{495}(4), 4845--4859.
\url{https://doi.org/10.1093/mnras/staa1530}

\bibitem[\citeproctext]{ref-de2026cosmocube}
de Lera Acedo, E., Bacon, D., Grainger, W., Anstey, D., Bevins, H.,
King, L., Artuc, K., Al-Badri, M., Pearson, C., Balaji, S., \& others.
(2026). The CosmoCube lunar mission for probing the dark ages and cosmic
dawn via 21-cm cosmology. \emph{Nature Astronomy}, \emph{10}(8),
1097--1106.

\bibitem[\citeproctext]{ref-Acedo2022REACH}
de Lera Acedo, E., de Villiers, D. I. L., Razavi-Ghods, N., Handley, W.,
Fialkov, A., Magro, A., Anstey, D., Bevins, H. T. J., Chiello, R.,
Cumner, J., Josaitis, A. T., Roque, I. L. V., Sims, P. H., Scheutwinkel,
K. H., Alexander, P., Bernardi, G., Carey, S., Cavillot, J., Croukamp,
W., \ldots{} Zarb-Adami, K. (2022). {The REACH radiometer for detecting
the 21-cm hydrogen signal from redshift z {\(\approx\)} 7.5-28}.
\emph{Nature Astronomy}, \emph{6}, 984--998.
\url{https://doi.org/10.1038/s41550-022-01709-9}

\bibitem[\citeproctext]{ref-Dhandha2025JWST21cm1}
Dhandha, J., Fialkov, A., Gessey-Jones, T., Bevins, H. T. J., Tacchella,
S., Pochinda, S., de Lera Acedo, E., Singh, S., \& Barkana, R. (2025a).
{Exploiting synergies between JWST and cosmic 21-cm observations to
uncover star formation in the early Universe}. \emph{MNRAS},
\emph{542}(3), 2292--2322. \url{https://doi.org/10.1093/mnras/staf1359}

\bibitem[\citeproctext]{ref-Dhandha2025JWST21cm2}
Dhandha, J., Fialkov, A., Gessey-Jones, T., Bevins, H. T. J., Tacchella,
S., Pochinda, S., de Lera Acedo, E., Singh, S., \& Barkana, R. (2025b).
{Narrowing the discovery space of the cosmological 21-cm signal using
multi-wavelength constraints}. \emph{MNRAS}, \emph{544}(2), 1608--1626.
\url{https://doi.org/10.1093/mnras/staf1736}

\bibitem[\citeproctext]{ref-DorigoJones2024LSTM21cm}
Dorigo Jones, J., Bahauddin, S. M., Rapetti, D., Mirocha, J., \& Burns,
J. O. (2024). {21CMLSTM: A Fast Memory-based Emulator of the Global 21
cm Signal with Unprecedented Accuracy}. \emph{ApJ}, \emph{977}(1), 19.
\url{https://doi.org/10.3847/1538-4357/ad8b20}

\bibitem[\citeproctext]{ref-DorigoJones2025KAN21cm}
Dorigo Jones, J., Reyes, B., Rapetti, D., Bahauddin, S. M., Burns, J.
O., \& Barker, D. W. (2025). {The Wrath of KAN: Enabling Fast, Accurate,
and Transparent Emulation of the Global 21 cm Cosmology Signal}.
\emph{ApJ}, \emph{991}(2), 152.
\url{https://doi.org/10.3847/1538-4357/adfc49}

\bibitem[\citeproctext]{ref-Fialkov2012RelativeMotion}
Fialkov, A., Barkana, R., Tseliakhovich, D., \& Hirata, C. M. (2012).
{Impact of the relative motion between the dark matter and baryons on
the first stars: semi-analytical modelling}. \emph{MNRAS},
\emph{424}(2), 1335--1345.
\url{https://doi.org/10.1111/j.1365-2966.2012.21318.x}

\bibitem[\citeproctext]{ref-Hirling2024pyc2ray}
Hirling, P., Bianco, M., Giri, S. K., Iliev, I. T., Mellema, G., \&
Kneib, J.-P. (2024). {pyC2Ray: A flexible and GPU-accelerated radiative
transfer framework for simulating the cosmic epoch of reionization}.
\emph{Astronomy and Computing}, \emph{48}, 100861.
\url{https://doi.org/10.1016/j.ascom.2024.100861}

\bibitem[\citeproctext]{ref-Kidger2022NeuralDE}
Kidger, P. (2022). {On Neural Differential Equations}. \emph{arXiv
e-Prints}, arXiv:2202.02435.
\url{https://doi.org/10.48550/arXiv.2202.02435}

\bibitem[\citeproctext]{ref-Lee2020HYREC2}
Lee, N., \& Ali-Ha\"{\i}moud, Y. (2020). {HYREC-2: A highly accurate
sub-millisecond recombination code}. \emph{PRD}, \emph{102}(8), 083517.
\url{https://doi.org/10.1103/PhysRevD.102.083517}

\bibitem[\citeproctext]{ref-Mellema2006C2ray}
Mellema, G., Iliev, I. T., Alvarez, M. A., \& Shapiro, P. R. (2006). {C
\(^{2}\)-ray: A new method for photon-conserving transport of ionizing
radiation}. \emph{New Astronomy}, \emph{11}(5), 374--395.
\url{https://doi.org/10.1016/j.newast.2005.09.004}

\bibitem[\citeproctext]{ref-Mesinger2011FAST21cm}
Mesinger, A., Furlanetto, S., \& Cen, R. (2011). {21CMFAST: a fast,
seminumerical simulation of the high-redshift 21-cm signal}.
\emph{MNRAS}, \emph{411}(2), 955--972.
\url{https://doi.org/10.1111/j.1365-2966.2010.17731.x}

\bibitem[\citeproctext]{ref-Mirocha2014ARES}
Mirocha, J. (2014). {Decoding the X-ray properties of pre-reionization
era sources}. \emph{MNRAS}, \emph{443}(2), 1211--1223.
\url{https://doi.org/10.1093/mnras/stu1193}

\bibitem[\citeproctext]{ref-Mittal2026ECHO21}
Mittal, S., Kulkarni, G., \& Sims, P. (2026). {ECHO21: a tool for
modelling global 21-cm signal from dark ages to reionization}. \emph{RAS
Techniques and Instruments}, \emph{5}, rzag001.
\url{https://doi.org/10.1093/rasti/rzag001}

\bibitem[\citeproctext]{ref-Munoz2023Zeus21}
Muñoz, J. B. (2023). {An effective model for the cosmic-dawn 21-cm
signal}. \emph{MNRAS}, \emph{523}(2), 2587--2607.
\url{https://doi.org/10.1093/mnras/stad1512}

\bibitem[\citeproctext]{ref-Murray2020FAST21cm}
Murray, S., Greig, B., Mesinger, A., Muñoz, J., Qin, Y., Park, J., \&
Watkinson, C. (2020). {21cmFAST v3: A Python-integrated C code for
generating 3D realizations of the cosmic 21cm signal.} \emph{The Journal
of Open Source Software}, \emph{5}(54), 2582.
\url{https://doi.org/10.21105/joss.02582}

\bibitem[\citeproctext]{ref-Pochinda2024PopIII}
Pochinda, S., Gessey-Jones, T., Bevins, H. T. J., Fialkov, A.,
Heimersheim, S., Abril-Cabezas, I., de Lera Acedo, E., Singh, S.,
Sikder, S., \& Barkana, R. (2024). {Constraining the properties of
Population III galaxies with multiwavelength observations}.
\emph{MNRAS}, \emph{531}(1), 1113--1132.
\url{https://doi.org/10.1093/mnras/stae1185}

\bibitem[\citeproctext]{ref-Reis2021LyAlpha21cm}
Reis, I., Fialkov, A., \& Barkana, R. (2021). {The subtlety of Ly
{\(\alpha\)} photons: changing the expected range of the 21-cm signal}.
\emph{MNRAS}, \emph{506}(4), 5479--5493.
\url{https://doi.org/10.1093/mnras/stab2089}

\bibitem[\citeproctext]{ref-Singh2022SARAS}
Singh, S., Jishnu, N. T., Subrahmanyan, R., Udaya Shankar, N., Girish,
B. S., Raghunathan, A., Somashekar, R., Srivani, K. S., \&
Sathyanarayana Rao, M. (2022). {On the detection of a cosmic dawn signal
in the radio background}. \emph{Nature Astronomy}, \emph{6}, 607--617.
\url{https://doi.org/10.1038/s41550-022-01610-5}

\bibitem[\citeproctext]{ref-SpurioMancini2026GRADSOLVE}
Spurio Mancini, A. (2026). {GRADSOLVE: fast exact gradients for ODE
ensembles on GPUs}. \emph{arXiv e-Prints}, arXiv:2609.02876.
\url{https://doi.org/10.48550/arXiv.2609.02876}

\bibitem[\citeproctext]{ref-Tutt2026Beams}
Tutt, J. L., Anstey, D. J., Cumner, J., Bevins, H. T. J., Cavillot, J.,
\& de Lera Acedo, E. (2026). {Towards end-to-end Bayesian forward models
in global 21-cm cosmology: surrogate modelling and marginalisation of
beam uncertainty}. \emph{arXiv e-Prints}, arXiv:2608.18962.
\url{https://doi.org/10.48550/arXiv.2608.18962}

\bibitem[\citeproctext]{ref-Tutt2026GPU21cm}
Tutt, J. L., Sims, P. H., Pattison, J. H. N., Anstey, D. J., Leeney, S.
A. K., \& de Lera Acedo, E. (2026). {Optimizing foreground modelling for
global 21-cm cosmology with GPU-accelerated nested sampling}.
\emph{MNRAS}, \emph{550}(1), stag1101.
\url{https://doi.org/10.1093/mnras/stag1101}

\bibitem[\citeproctext]{ref-Visbal2012FirstStars}
Visbal, E., Barkana, R., Fialkov, A., Tseliakhovich, D., \& Hirata, C.
M. (2012). {The signature of the first stars in atomic hydrogen at
redshift 20}. \emph{Nature}, \emph{487}(7405), 70--73.
\url{https://doi.org/10.1038/nature11177}

\end{CSLReferences}
\end{document}